\documentclass[pra,twocolumn,footinbib]{revtex4}
\usepackage[latin1]{inputenc}
\usepackage{amsmath,amssymb,epsfig}

\newcommand{\assign}{:=}
\newcommand{\mathd}{\mathrm{d}}
\newcommand{\mathe}{\mathrm{e}}
\newcommand{\tmem}[1]{{\em #1\/}}
\newcommand{\tmmathbf}[1]{\ensuremath{\boldsymbol{#1}}}
\newcommand{\tmop}[1]{\ensuremath{\operatorname{#1}}}

\newcommand{\um}{-}
\newcommand{\bignone}{\,}
\newcommand{\bigintlim}{\int}

\begin{document}

\title{Thermal limitation of far-field matter-wave interference}\author{Klaus
Hornberger}
\affiliation{Arnold Sommerfeld Center for Theoretical Physics,
  Ludwig-Maximilians-Universität München, Theresienstr.~37,
  80333 Munich, Germany}

\begin{abstract}
  We assess the effect of the heat radiation emitted by mesoscopic particles on
  their ability to show interference in a double slit arrangement. The
  analysis is based on a {\em stationary}, phase-space based description of
  matter wave interference in the presence of momentum-exchange mediated
  decoherence. 
\end{abstract}
\maketitle
\section{Introduction}

A recent demonstration of the wave nature of large-sized molecules with a high
internal temperature {\cite{Hackermuller2003a}} poses the question on the
ultimate physical limitations for the observation of matter wave interference
with increasingly macroscopic particles {\cite{Schlosshauer2006a}}. The
present article focuses on one of the basic effects known to destroy the
coherence of matter waves, namely the heat radiation emitted by the particles
themselves. We take these objects to be sufficiently complex so that their
internal state calls for a thermodynamic description and obtain the general
scaling behavior as a function of their temperature, heat capacity, and
effective surface area.

As a second motive, we note that the loss of coherence in a two-path
matter wave interferometer has been the subject of a number of recent
studies, discussing various models and mechanisms
{\cite{FacchiNonsense,Viale2003a,Levinson2004a,Sanz2005a,Lombardo2005a,Machnikowski2005b1,Qureshi2006a1}}.
All of those treatments employ a {\tmem{temporal}} description of the
decohering dynamics, usually by approximately solving a master
equation or a corresponding stochastic differential equation. Here we
present an alternative way of dealing with decoherence of matter wave
beams, which is based on describing the rate of decoherence events and
their effect separately. Using the Wigner-Weyl formulation of quantum
mechanics one finds how the {\tmem{stationary}} scattering state of
the particle beam gets affected by momentum-exchange mediated
decoherence. This way a clear and transparent description of the
underlying physics is obtained.

We consider the most basic situation of far-field interference where a
stationary beam of matter waves passes a double slit arrangement,
although the treatment is easily generalized to multi-slit gratings.
The application to the case of heat radiation yields simple, yet
experimentally relevant formulas for the expected loss of the
interference contrast, and permits to incorporate and discuss
important issues such as the effects of cooling and and of deviations
from the ideal blackbody spectrum.

The experimental studies of the impact of light emission on matter
waves started with the scattering of laser beams off interfering atoms
{\cite{Pfau1994a,Chapman1995a,Kokorowski2001a}}. More recently, the
decohering effect of the continuous, Planck-like heat radiation
emitted by fullerenes was observed, in agreement with decoherence
theory, in the near-field regime of Talbot-Lau interference
{\cite{Hackermuller2004a,Hornberger2005a}}.  Theoretical discussions
of the general effect of heat radiation serving to ``localize'' a
spatially extended quantum state into a mixture of more local states
can be found in
{\cite{Joos1985a,Zurek1991a,Tegmark1993a,Durr1999a,Alicki2002a,Joos2003a}},
though not in the context of interferometry.

The structure of the article is as follows. Section 2 starts by collecting
the required equations to treat general momentum-exchange mediated
decoherence. Since the approach was already used in {\cite{Hornberger2004a}}
to treat the related case of near-field interference we keep the presentation
short. Section~3 proceeds to calculate the double slit far-field interference
pattern in the presence of decoherence. Based on this we discuss the
visibility reduction due to heat radiation, i.e., the thermal limitation for
interference, with mesoscopic particles in Sect.~4. Quantitative estimates are
presented for carbonaceous aerosols, and compared to the case of fullerene
molecules. The article concludes with with Sect.~5.

\section{Decoherence mediated by momentum transfer}

The ability of a particle to show interference may be reduced due the
interaction and entanglement with unobserved degrees of freedom, such as
collisions with particles from a background gas or the absorption and emission
of photons. It is often justified, and consistent with the Markov assumption,
to model this loss of coherence as due to individual events, which occur
probabilistically and independently of each other. The situation is
particularly simple if the motional state of the particle $\hat{\varrho}'$,
obtained by tracing over the unobserved degrees of freedom, depends only on
the momentum exchange during the event. In position representation, the
off-diagonal elements are then merely reduced by a complex factor $\eta$, with
$\left| \eta \right| \leqslant 1$. The reduction depends only on the
corresponding distance,
\begin{eqnarray}
  \langle \tmmathbf{R}_1 | \hat{\varrho}' |\tmmathbf{R}_2 \rangle & = &
  \langle \tmmathbf{R}_1 | \hat{\varrho} |\tmmathbf{R}_2 \rangle\, 
\eta \left(
    \tmmathbf{R}_1 -\tmmathbf{R}_2 \right),  \label{eq:eta3d}
\end{eqnarray}
while the diagonal elements are unaffected, $\eta \left( 0 \right) = 1$. This
loss of coherence can be attributed to the amount of which-way information
revealed to the environment after the interaction {\cite{Arndt2005a}}. At the
same time, the ``decoherence function'' $\eta$ is given by the Fourier
transform of the probability density for the momentum exchange, $\eta \left(
\Delta \tmmathbf{R} \right) = \eta^{\ast} \left( \um \Delta \tmmathbf{R}
\right) = {\int \bignone \mathd^3 \tmmathbf{Q} \exp \left( i \Delta
\tmmathbf{R} \cdot \tmmathbf{Q}/ \hbar \right) \overline{\eta} \left(
\tmmathbf{Q} \right)},$ which is manifest in the operator representation of
(\ref{eq:eta3d}),
\begin{eqnarray}
  \hat{\varrho}' & = & \int \bignone \mathd^3 \tmmathbf{Q}\, \bar{\eta} \left(
  \tmmathbf{Q} \right) \mathe^{i \widehat{\mathsf{R}} \cdot \tmmathbf{Q}/
  \hbar} \bignone \hat{\varrho}\bignone  \mathe^{- i \widehat{\mathsf{R}} \cdot
  \tmmathbf{Q}/ \hbar} .  \label{eq:rhoprime}
\end{eqnarray}
In fact, in the more general framework of translation-invariant and completely
positive master equations the function $\eta$ can be related to the
characteristic function of the relevant Poissonian part of the corresponding
Lévy process {\cite{Vacchini2005a}}. For later reference we note that in terms
of the Wigner function the state change (\ref{eq:rhoprime}) reads
\begin{eqnarray}
  W' \left( \tmmathbf{R},\tmmathbf{P} \right) & = & \int \mathd^3 \tmmathbf{Q}
\,  \bar{\eta} \left( \tmmathbf{Q} \right) W \left(
  \tmmathbf{R},\tmmathbf{P}-\tmmathbf{Q} \right) .  \label{eq:Wchange}
\end{eqnarray}
The time evolution equation for the resulting decohering motion is obtained
immediately from (\ref{eq:eta3d}) if one assumes that the decoherence events
occur with a rate $\gamma \left( t \right)$. It has the Markovian form
$\partial_t \hat{\varrho} = \left( i \hbar \right)^{- 1} [
\widehat{\mathsf{H}}, \hat{\varrho} ] + \mathcal{L}_t \hat{\varrho}$, with the
incoherent part given by
\begin{eqnarray}
  \langle \tmmathbf{R}_1 | \mathcal{L}_t \hat{\varrho} |\tmmathbf{R}_2 \rangle
  & = & - \gamma \left( t \right)  \left[ 1 - \eta \left( \tmmathbf{R}_1
  -\tmmathbf{R}_2 \right) \right] \langle \tmmathbf{R}_1 | \hat{\varrho}
  |\tmmathbf{R}_2 \rangle . \nonumber\\
  &  &  \label{eq:me}
\end{eqnarray}
The process of decoherence by the emission of heat radiation fits into the
present framework provided the particle beam is unpolarized and the density of
the electro-magnetic field modes can be taken constant in space, i.e., if the
modification of the emission rate due to nearby surfaces can be neglected. If
the walls of the apparatus are located far away compared to the photon wave
length the sole effect of a photon emission on the motional state of the
particle is an isotropic momentum kick (the internal degrees of freedom remain
in a separable state). One obtains the corresponding decoherence function by
expressing the probability density for a photon to be emitted at (angular)
frequency $\omega$ in terms of the temperature dependent spectral photon
emission rate {$R_{\omega} \left( \omega ; T \right)$}. The result is
\begin{eqnarray}
  \eta \left( \tmmathbf{R} \right) & = & \frac{1}{R_{\tmop{tot}}}
  \int_0^{\infty} \mathd \omega R_{\omega} \left( \omega ; T \right)
  \tmop{sinc} \left(  \frac{\omega}{c}  \left| \tmmathbf{R} \right| \right) 
  \label{eq:etarad}
\end{eqnarray}
with total emission rate $R_{\tmop{tot}} = \int_0^{\infty} \mathd \omega
R_{\omega} \bignone \left( \omega ; T \right)$ and $\tmop{sinc} \left( x
\right) \equiv \sin \left( x \right) / x.$ Putting aside this particular form
we keep $\eta$ unspecified in the following section, thereby including all
other (possibly anisotropic) momentum-transfer mediated decoherence
mechanisms.

\section{Decoherence in double slit interference}

\subsection{Phase space description of double slit interference}

We proceed to formulate the double-slit interference of matter waves in the
phase space representation (e.g., {\cite{Ozorio1998a,Schleich2001a}}). This
will permit us to incorporate the effect of decoherence in the subsequent
section.

Consider the usual interferometric situation with a well-collimated,
stationary beam of particles directed in the positive $z$-direction. Noting
that the stationarity implies the absence of coherences between different
longitudinal momenta {\cite{Rubenstein1999b}}, it is represented by a
(non-normalizable) state where the longitudinal and transverse motion are
initially separable,
\begin{eqnarray}
  \hat{\varrho}_{\tmop{in}} & = & \int \mathd p_z  \bignone g \left( p_z
  \right)  \hat{\rho}  \left( p_z \right) \otimes |p_z \rangle \langle p_z |, 
\end{eqnarray}
with $g \left( p_z \right)$ the longitudinal momentum distribution function and
$\hat{\rho}$ a transverse state. Since this is a convex sum we can at first
take the incoming state of the beam to have a well-defined longitudinal
velocity $p_z / m$, and discuss the effect of a finite longitudinal coherence
length later. Lower-case vectors are used to denote coordinates in the
transverse plane, e.g., $\tmmathbf{r} \equiv \left( x, y, 0 \right)$.
Accordingly, the Wigner function of the transverse motion is given by
\begin{eqnarray}
  w \left( \tmmathbf{r},\tmmathbf{p} \right) & = & \frac{1}{\left( 2 \pi \hbar
  \right)^2} \int \mathd^2 \tmmathbf{q}\, \mathe^{- i\tmmathbf{r} \cdot
  \tmmathbf{q}/ \hbar} \bignone  \\
  &  & \phantom{\frac{1}{\left( 2 \pi \hbar \right)^2} \int} \times \langle
  \tmmathbf{p}- \frac{\tmmathbf{q}}{2} | \hat{\rho} \left( p_z \right)
  |\tmmathbf{p}+ \frac{\tmmathbf{q}}{2} \rangle . \nonumber
\end{eqnarray}
Now, if the passage through a single pinhole rendered the transverse motion in
the pure state $| \psi_{\text{s}} \rangle$ a double slit arrangement with
identical pinholes would produce the state
\begin{eqnarray}
  | \psi_0 \rangle & = & \frac{1}{\sqrt{2}}  \left( e^{i \hat{\mathsf{p}}
  \cdot \tmmathbf{d}/ 2 \hbar} + e^{- i \hat{\mathsf{p}} \cdot
  \tmmathbf{d}/ 2 \hbar} \right) | \psi_{\text{s}} \rangle,  \label{eq:psi0}
\end{eqnarray}
provided there is full transverse coherence over the double slit separation
$\tmmathbf{d}$. As implied by the normalization, we take the translated
pinhole states to be non-overlapping, $\langle \psi_{\text{s}} | e^{- i
\hat{\mathsf{p}} \cdot \tmmathbf{d}/ \hbar} | \psi_{\text{s}} \rangle =
0$. Allowing for arbitrary pinhole states $\hat{\rho}_{\text{s}}$ the relation
(\ref{eq:psi0}) reads $\hat{\rho}_0 = 2 \cos \left( \hat{\mathsf{p}} \cdot
\tmmathbf{d}/ 2 \hbar \right)  \hat{\rho}_{\text{s}} \cos \left(
\hat{\mathsf{p}} \cdot \tmmathbf{d}/ 2 \hbar \right)$, and accordingly we
have in phase space representation:
\begin{eqnarray}
  w^{\left( 0 \right)} \left( \tmmathbf{r},\tmmathbf{p} \right) & = &
  \frac{1}{2}  w^{\left( \text{s} \right)} \left( \tmmathbf{r}+
  \frac{\tmmathbf{d}}{2},\tmmathbf{p} \right) + \frac{1}{2} w^{\left( \text{s}
  \right)} \left( \tmmathbf{r}- \frac{\tmmathbf{d}}{2},\tmmathbf{p} \right) .
  \nonumber\\
  &  & + \cos \left( \frac{\tmmathbf{p} \cdot \tmmathbf{d}}{\hbar} \right)
  w^{\left( \text{s} \right)} \left( \tmmathbf{r},\tmmathbf{p} \right) . 
\end{eqnarray}
This implies that the probability density of the transverse momentum is
related to the corresponding single slit distribution $w_p^{\left( \text{s}
\right)} \left( \tmmathbf{p} \right) \assign \int \mathd^2 \tmmathbf{r}
\bignone w^{\left( s \right)} \left( \tmmathbf{r},\tmmathbf{p} \right)$ by
\begin{eqnarray}
  w_p^{\left( 0 \right)} \left( \tmmathbf{p} \right) & \assign & \int \mathd^2
  \tmmathbf{r} \bignone w^{\left( 0 \right)} \left( \tmmathbf{r},\tmmathbf{p}
  \right) = \nonumber\\
  & = & \left[ 1 + \cos \left( \frac{\tmmathbf{p} \cdot \tmmathbf{d}}{\hbar}
  \right) \right] w_p^{\left( \text{s} \right)} \left( \tmmathbf{p} \right) . 
\end{eqnarray}
The great advantage of the phase space representation shows up in the the free
evolution from the double slit, at $z = 0$ to the screen located at $z = L$,
since the corresponding states are related by the free shearing transformation
(e.g.~{\cite{Hornberger2004a}})
\begin{eqnarray}
  w^{\left( L \right)} \left( \tmmathbf{r},\tmmathbf{p} \right) = w^{\left( 0
  \right)} \left( \tmmathbf{r}- L \frac{\tmmathbf{p}}{p_z},\tmmathbf{p}
  \right) . &  &  \label{eq:Wfp}
\end{eqnarray}
With this one arrives at the familiar result that the interference pattern in
the far field is given by the momentum distribution of the particle state
after it passed the grating,
\begin{eqnarray}
  w^{\left( L \right)}_r \left( \tmmathbf{r} \right) & \assign & \int \mathd^2
  \tmmathbf{p}\,w^{\left( L \right)} \left( \tmmathbf{r},\tmmathbf{p} \right)
  \nonumber\\
  & = & \frac{p_z^2}{L^2} \int \mathd^2 \tmmathbf{r}_0 \,w^{\left( 0 \right)}
  \left( \tmmathbf{r}_0, p_z  \frac{\tmmathbf{r}-\tmmathbf{r}_0}{L} \right) 
  \nonumber\\
  & \cong & \frac{p_z^2}{L^2} \int \mathd^2 \tmmathbf{r}_0\, w^{\left( 0
  \right)} \left( \tmmathbf{r}_0, p_z  \frac{\tmmathbf{r}}{L} \right) 
  \hspace{1em} \text{for $\left| \tmmathbf{r} \right| \gg \left| \tmmathbf{d}
  \right|$} \nonumber\\
  & = & \frac{p_z^2}{L^2} w_p^{\left( 0 \right)} \left( p_z 
  \frac{\tmmathbf{r}}{L} \right) .  \label{eq:farfield}
\end{eqnarray}
As the only approximation we used here that $w^{\left( 0 \right)} \left(
\cdot,\tmmathbf{p}_0 \right)$ is localized in the small pinhole areas so that
one can neglect the $p_z \tmmathbf{r}_0 / L$ term for $\text{for $\left|
\tmmathbf{r} \right| \gg \left| \tmmathbf{d} \right|$}$. This is equivalent to
the far-field approximation of Fraunhofer diffraction. Insertion of the
momentum distribution after the double slit yields the far field pattern
\begin{eqnarray}
  w^{\left( L \right)}_r \left( \tmmathbf{r} \right) & = &  \frac{p_z^2}{L^2} 
  \left[ 1 + \cos \left( \frac{p_z}{L}  \frac{\tmmathbf{r} \cdot
  \tmmathbf{d}}{\hbar}  \right) \right] w_p^{\left( \text{s} \right)} \left(
  p_z  \frac{\tmmathbf{r}}{L} \right),  \label{eq:wrds}
\end{eqnarray}
and it shows the expected modulation of the single slit momentum distribution
by interference fringes with a period of $2 \pi \hbar L / \left( p_z \left|
\tmmathbf{d} \right| \right) = \lambda_{\text{dB}} L / \left| \tmmathbf{d}
\right|$. It is common to attribute a visibility of $\mathcal{V} = 1$ to this
pattern.

In experiments, one typically uses detectors, which measure the
{\tmem{intensity}} of the interference pattern (the number of particles per
unit area and time). In addition, one should take into account that molecular
beams cannot be prepared in a monochromatic state, but have a finite
longitudinal coherence.  By allowing for a distribution of the longitudinal
momenta $g \left( p_z \right)$ the expected intensity is proportional
to {\footnote{In principle, $w^{\left( L \right)}_r \left( \tmmathbf{r}
\right)$ may depend on $p_z$ via $w_p^{\left( \text{s} \right)} \left(
\tmmathbf{p} \right)$, e.g. if the particle-grating interaction depends on the
time of traversal through the slit. This effect is incorporated in the
formalism, but not made explicit here due its relative unimportance for
far-field diffraction.}}
\begin{eqnarray}
  I \left( \tmmathbf{r} \right) & = & \int \mathd p_z \,g \left( p_z \right)
  \bignone  \frac{p_z}{m} w^{\left( L \right)}_r \left( \tmmathbf{r} \right) .
  \label{eq:I1}
\end{eqnarray}
However, to keep the formulation simple we take a fixed value of $p_z$ in the
following, and note that the general case is simply recovered by weighting the
resulting {\tmem{intensities}} with $g \left( p_z \right)$. In any case, for
the beams used in molecular interferometry we can safely assume the
contributing $p_z$ to be much larger than the possible longitudinal momentum
transfers, so that the change of the velocity due to photon emission can be
neglected in the following section, where we proceed to account for
decoherence.

\subsection{Incorporating decoherence}

At first sight, the obvious way to incorporate the effect of decoherence would
be to solve the corresponding master equation (\ref{eq:me}). As shown
in 
Sect.~VI. of Ref.~{\cite{Hornberger2004a}}, a systematic scheme can indeed be set
up which yields successive approximations of the corresponding time evolution
in a closed form. However, by insisting on a dynamic description to treat an
inherently stationary problem one loses much transparency in the formulation.
This renders a stationary treatment much more appropriate, as will be clear in
the following.

Let us first consider the impact of a single decoherence event occurring at
some longitudinal position $z$. The interference pattern conditioned on this
event is obtained by propagating the state to position $z$ with
(\ref{eq:Wfp}), performing the decohering convolution (\ref{eq:Wchange}), and
a further propagation to the final screen position $L$. It follows that the
resulting intensity pattern is given by
\begin{eqnarray}
  I' \left( z ;\tmmathbf{r} \right) & = & \frac{p_z}{m} \int \mathd^2
  \tmmathbf{q} \bignone \mathd q_z \int \mathd^2 \tmmathbf{p} \bignone 
  \bar{\eta} \left( \tmmathbf{q}+ q_z \tmmathbf{e}_z \right) \nonumber\\
  &  & \times w^{\left( 0 \right)} \left( \tmmathbf{r}- L
    \frac{\tmmathbf{p}}{p_z} + z
    \frac{\tmmathbf{q}}{p_z},\tmmathbf{p}-\tmmathbf{q} \right) \nonumber\\
  & \cong & \bignone  \frac{p_z^3}{mL^2} \int \mathd^2 \tmmathbf{q}\,w^{\left(
      0 \right)}_p \left( p_z \frac{\tmmathbf{r}}{L} - \frac{L - z}{L}
    \tmmathbf{q} \right) \nonumber\\
  &  & \times \int \mathd q_z \bignone  \bar{\eta} \left( \tmmathbf{q}+ q_z
    \tmmathbf{e}_z \right) \nonumber\\
  & = & \int \mathd^2 \tmmathbf{q}I \left( \tmmathbf{r}- \frac{L - z}{p_z}
    \tmmathbf{q} \right) \int \mathd q_z \bignone  \bar{\eta} \left(
    \tmmathbf{q}+ q_z \tmmathbf{e}_z \right) .  
\nonumber\\
&&\label{eq:Iprime}
\end{eqnarray}
Here we performed the same far-field approximation as above in
(\ref{eq:farfield}), took into account that only $\left| q_z \right| \ll p_z$
contribute in $\bar{\eta}$, and related the result to the unperturbed
intensity pattern (\ref{eq:I1}). In the same fashion, one finds that
additional decoherence events are described by further convolutions with
$\bar{\eta}$.

The trick is now to consider the differential equation which governs the
change of the interference pattern as one increases the longitudinal interval
where decoherence events may occur. This is straightforward since the patterns
are related by a convolution. After taking the Fourier transform of the
interference pattern, $\bar{I} \left( \tmmathbf{q} \right) = \left( 2 \pi
\hbar \right)^{- 2} {\int \mathd^2 \tmmathbf{r} \exp \left( - i\tmmathbf{r}
\cdot \tmmathbf{q}/ \hbar \right) I \left( \tmmathbf{r} \right)} \bignone$,
the relation (\ref{eq:Iprime}) turns into a multiplication with the
decoherence function,
\begin{eqnarray}
  \bar{I}' \left( z ;\tmmathbf{q} \right) & = & \bar{I} \left( \tmmathbf{q}
  \right) \eta \left(  \frac{z - L}{p_z} \tmmathbf{q} \right) . 
  \label{eq:Iqprime}
\end{eqnarray}
We denote by $\Gamma_z \left( z \right) \mathd z$ the (possibly position
dependent) average number of decoherence events occurring in $\left( z, z +
\mathd z \right)$. The differential change of the pattern is then determined
by
\begin{eqnarray}
  \frac{\mathd}{\mathd z} \bar{I}_{\tmop{dec}} \left( z ;\tmmathbf{q} \right)
  & = & \Gamma_z \left( z \right) \left[  \bar{I}_{\tmop{dec}} \left( z
  ;\tmmathbf{q} \right) \eta \left( \tmmathbf{q} \frac{z - L}{p_z} \right) 
  \label{eq:de} \right.\\
  &  & \left. \phantom{\Gamma_z \left( z \right) \left[ \right.} -
  \bar{I}_{\tmop{dec}} \left( z ;\tmmathbf{q} \right) \right] . \nonumber
\end{eqnarray}
The integration from $z = 0$ to $z = L$ yields 
\begin{eqnarray}
  \bar{I}_{\tmop{dec}} \left( \tmmathbf{q} \right) & = & \exp \left[ -
  \int_0^L \bignone \mathd z \Gamma_z \left( z \right) \left( 1 - \eta \left(
  \tmmathbf{q} \frac{z - L}{p_z} \right) \right) \bignone \right] 
 \nonumber\\
  &  &\times\bar{I}  \left( \tmmathbf{q} \right) .  \label{eq:Idecq}
\end{eqnarray}
After a final Fourier transform we obtain the general result
\begin{eqnarray}
  I_{\tmop{dec}} \left( \tmmathbf{r}) \right. & = & \int \mathd^2
  \tmmathbf{s}\,h \left( \tmmathbf{s} \right) I \left( \tmmathbf{r}-\tmmathbf{s}
  \right)  \bignone .  \label{eq:Idec1}
\end{eqnarray}
It shows that the entire effect of decoherence on the interference pattern is
described by a convolution with the real kernel
\begin{eqnarray}
  h \left( \tmmathbf{s} \right) & \assign & \frac{1}{\left( 2 \pi \hbar
  \right)^2} \int \mathd^2 \tmmathbf{q} \,\mathe^{i\tmmathbf{q} \cdot
  \tmmathbf{s}/ \hbar}   \label{eq:hdef}\\
  &  &  \times \exp
  \left[ - \int_0^L \bignone \mathd z\, \Gamma_z \left( z \right) \left( 1 -
  \eta \left( \tmmathbf{q} \frac{z - L}{p_z} \right) \right) \bignone \right]
  . \nonumber
\end{eqnarray}
It is determined by the decoherence function from (\ref{eq:eta3d}) and the
(possibly time-dependent) rate of decoherence events $\gamma \left( t
\right)$, related to $\Gamma_z$ by $\Gamma_z \left( z \right) v_z = \gamma
\left( z / v_z \right)$ with $v_z = p_z / m$. (The fact that (\ref{eq:Idec1})
is positive for all intensity patterns $I \left( \tmmathbf{r} \right)$ follows
with Bochner's theorem by noting that the exponential function in
(\ref{eq:Idecq}) is of positive type since $\eta$ is of positive type.) It
should be emphasized that multiple decoherence events are treated correctly in
this description, provided they are statistically independent, since
(\ref{eq:de}) is a linear equation.

Note that the result (\ref{eq:Idec1}), (\ref{eq:hdef}) is valid for any
initial momentum distribution $w_p^{\left( 0 \right)}$ thus covering arbitrary
multi-slit interference arrangements. If we go back to the case of a double
slit (\ref{eq:wrds}) an approximate, but very intuitive form is obtained if
the width of $w^{\left( \text{s} \right)}_p$ is so large that one can pull the
envelope of the interference pattern out of the $\tmmathbf{s}$-integration. If
we also take $\eta$ to be isotropic (and therefore real) the interference
pattern reads in this case
\begin{eqnarray}
  I_{\tmop{dec}} \left( \tmmathbf{r}) \right. & \cong & \frac{v_z p_z^2}{L^2} 
  \left[ 1 +\mathcal{V} \cos \left( \frac{p_z}{\hbar}  \frac{\tmmathbf{r}
  \cdot \tmmathbf{d}}{L} \right) \right] w^{\left( \text{s} \right)}_p \left(
  p_z  \frac{\tmmathbf{r}}{L} \right) . \nonumber\\
  &  &  \label{eq:Idec2}
\end{eqnarray}
It differs from the unperturbed pattern, see (\ref{eq:wrds}), by the factor
\begin{eqnarray}
  \mathcal{V} & = & \exp \left[ - \frac{1}{v_z} \int_0^L \bignone \mathd z
 \, \gamma \left( \frac{z}{v_z} \right)  \left( 1 - \eta \left( \tmmathbf{d}
  \frac{L - z}{L} \right) \right) \bignone \right], \nonumber\\
  &  &  \label{eq:Vred}
\end{eqnarray}
which reduces the oscillating term. Although there is no unique definition for
the visibility of a non-periodic signal like (\ref{eq:Idec2}), one may call
this the {\tmem{visibility reduction}} due to decoherence.

Equation (\ref{eq:Vred}) has a particularly intuitive form since it involves
the separation of the two interfering paths in the argument of the decoherence
function. Close to the double-slit, at $z = 0$, the decoherence function
enters with the full slit separation $\left| \tmmathbf{d} \right|$. As the
particle approaches the screen the two paths get closer, requiring an
increasingly large ``resolving power'' of the decoherence events if they are
to contribute to the visibility reduction by conveying some which way
information. At the screen, where $z = L$ and $\eta \left( 0 \right) = 1$ the
paths coalesce and decoherence has no effect. This dependence on the
longitudinal position has been observed in experiments where a laser beam was
scattered off interfering atoms
{\cite{Pfau1994a,Chapman1995a,Kokorowski2001a}}, albeit in a Mach-Zehnder
setup. It is also an important factor in the quantitative description of the
recent near-field experiment on decoherence by heat radiation
{\cite{Hackermuller2004a}}.

Note that in the case of {\tmem{near-field}} Talbot-Lau interference an
approximate expression for the visibility reduction has been found in Ref.
{\cite{Hornberger2004a}}. It is similar to (\ref{eq:Vred}), but the
denominator in the argument of of $\eta$ is replaced by (an integer fraction
of) the Talbot length $d^2 / \lambda_{\tmop{dB}}$. While this underlines the
strength of the phase space approach in incorporating decoherence in the
various regimes of interferometry, we note that the intermediate case, where
neither near-field nor far-field approximations hold, is more complicated.

\section{Thermal scaling behavior}

Based on the results of the preceding section we can now study at what
temperatures the wave nature of material objects ceases to be observable due
to their heat radiation. For a start, it should be noted that, strictly
speaking, one can apply (\ref{eq:Vred}) only if the thermal photons are
emitted in a Poisson process, which would require the particle to be
permanently in contact with a heat bath. The isolated, mesoscopic
particle-waves traversing the interferometer are typically {\tmem{not}} in
thermal equilibrium with the surrounding radiation field, and they are
characterized by the amount of energy stored in their many internal degrees of
freedom. One may express this internal energy in terms of the micro-canonical
temperature $T$ and the heat capacity $C_V$.

As a consequence of the absence of a proper heat bath each photon emission
reduces the micro-canonical temperature by $\hbar \omega / C_V$. This modifies
the emission probability of the subsequent photons and renders the process
non-Poissonian. However, in an approximate description the main consequences
of this this effect can be accounted for by an inhomogeneous Poisson process
with a time-dependent emission rate $\gamma \left( t \right) = R_{\tmop{tot}}
\left( T \left( t \right) \right)$ where $T \left( t \right)$ is determined by
the cooling equation $\partial_t T \left( t \right) = - \bigintlim \mathd
\omega \bignone \hbar \omega R_{\omega} \left( \omega ; T \left( t \right)
\right) / C_V$. One writes (\ref{eq:Vred}) as a time integration and inserts
(\ref{eq:etarad}) which cancels the $R_{\tmop{tot}}$ dependence. The resulting
visibility reduction due to heat radiation reads
\begin{eqnarray}
  \mathcal{V} & = & \exp \left[ - \int_0^{\tau} \bignone \mathd t
  \int_0^{\infty} \mathd \omega R_{\omega} \left( \omega ; T \left( t \right)
  \right)  \label{eq:Vth} \right.\\
  &  & \left. \phantom{\exp \left[ \right.} \times \left\{ 1 - \tmop{sinc}
  \left( \frac{\omega d}{c}  \left( 1 - \frac{t}{\tau} \right) \right)
  \right\} \right] \nonumber
\end{eqnarray}
with $\tau \assign L / v_z$ the time of flight and $d = \left| \tmmathbf{d}
\right|$ the slit separation. Likewise, the corresponding full intensity
pattern is obtained by replacing the exponential in the kernel (\ref{eq:hdef})
with the right hand side of (\ref{eq:Vth}), after the substitution $d
\rightarrow \left| \tmmathbf{q} \right| L / p_z$. As discussed above, the
calculation of this modified pattern is required if one has to account for a
finite coherence length.

As exemplified below, the influence of cooling can be quite significant in
realistic scenarios. For objects with a molecular structure it is even more
crucial to account for the full frequency dependence of the emission rate. The
reason is that the rate is strongly affected by the electronic excitation
spectrum of the molecule, which typically shows a gap at optical to near
infrared frequencies, leading to a much stronger temperature and frequency
dependence than expected for a blackbody radiator.

In the following we consider interfering objects which are still considerably
larger than molecules, so that the absorption cross section
$\sigma_{\tmop{abs}}$ can be taken frequency and temperature independent and
proportional to the surface area $\mathcal{A}$. Although this is unrealistic
for molecules, it is quite reasonable for submicrometer particles like aerosols
or large proteins, which display an essentially featureless absorption cross
section in the relevant frequency regime {\cite{Bohren1983a}}. In this case
one accounts for the deviations with respect to the macroscopic blackbody
spectrum by introducing the emissivity $\varepsilon$, which characterizes the
effective emission area{\footnote{For a spherical particle the effective area
is related to the absorption cross section $\sigma_{\tmop{abs}}$ by
$\mathcal{A}_{\varepsilon} = 4 \sigma_{\tmop{abs}}$.}}
$\mathcal{A}_{\varepsilon} \assign \varepsilon \mathcal{A}$. The spectral
photon emission rate is then given by {\cite{Hansen1998a}}
\begin{eqnarray}
  R_{\omega} \left( \omega, T \right) & = & \frac{\mathcal{A}_{\varepsilon}
  \omega^2}{\left( 2 \pi c \right)^2} \exp \left( - \frac{\hbar
  \omega}{k_{\text{B}} T} - \frac{k_B}{2 C_V} \left( \frac{\hbar
  \omega}{k_{\text{B}} T} \right)^2 \right) . \nonumber\\
  &  &  \label{eq:Rarea}
\end{eqnarray}
The statistical factor differs here from the usual Bose-Einstein distribution
function because the particle is not in thermal equilibrium with the colder,
surrounding radiation field, so that induced absorption plays no role. The
term involving the heat capacity is present since the emission reduces the
internal energy.

Solving the cooling equation with the emission spectrum (\ref{eq:Rarea}) one
finds that the particle temperature decreases like $T \left( t \right) = T_0 [
1 + T_0^3 / T^3_{\inf} \left( t \right) ]^{- 1 / 3}$. Here, the limiting
temperature at time $t$ is determined by $T_{\inf}^{- 3} \left( t \right) = 9
k_{\text{B}}^4  \mathcal{A}_{\varepsilon} tF \left( C_V / k_B \right) / \left(
2 \pi^2 c^2 \hbar^3 C_V \right)$ where the factor $F \left( x \right) = 1 - 10
x^{- 1} + 105 x^{- 2} + \mathcal{O} \left( x^{- 3} \right)$ accounts for the
$C_V$-dependence in (\ref{eq:Rarea}). Below, this result is used to give a
quantitative estimate for aerosols.

As a final step, we take the particle to be truly macroscopic in the sense
that the heat capacity is so large that the effect of cooling can be
neglected, $C_V / k_{\text{B}} \rightarrow \infty$. It follows from
(\ref{eq:Vth}) that the visibility then decays exponentially with the time of
flight, i.e., $\mathcal{V}= \exp \left( \um \tau / \tau_{\tmop{th}} \right)$.
The characteristic decoherence time $\tau_{\tmop{th}}$ is determined by a
spectral integration involving the sine integral $\tmop{Si}(x)$
{\cite{Abramowitz1965a}},
\begin{eqnarray}
  \tau_{\tmop{th}}^{- 1} & = & \int_0^{\infty} \mathd \omega R_\omega \left( \omega
  ; T \right) \left[ 1 - \frac{c}{\omega d} \tmop{Si} \left(  \frac{\omega
  d}{c} \right)  \right] .  \label{eq:tautherm1}
\end{eqnarray}
With the special choice (\ref{eq:Rarea}) for the emission rate the integral
can be done. One obtains the final result for the decoherence time
\begin{eqnarray}
  \tau_{\tmop{th}}^{- 1} & = & \frac{\mathcal{A}_{\varepsilon} c}{\left( 2 \pi
  \right)^2 d^3} f \left( \frac{k_{\text{B}} Td}{\hbar c} \right)  
  \label{eq:tautherm2}
\end{eqnarray}
with the scaling function
\begin{eqnarray}
  f \left( x \right) & = & 2 x^3 - \frac{x^3}{1 + x^2} - x^2 \arctan \left( x
  \right) . \nonumber
\end{eqnarray}
The expansion for small arguments renders the reciprocal decoherence time
proportional to the square of the slit separation $d$ and to the fifth power
of the temperature $T$, or Matsubara frequency,
\begin{eqnarray}
  \tau_{\tmop{th}}^{- 1} & = & \frac{1}{3 \pi^2} 
  \frac{\mathcal{A}_{\varepsilon} d^2}{c^4}  \left( \frac{k_{\text{B}}
  T}{\hbar} \right)^5 \times \left[ 1 + \mathcal{O} \left( \frac{k_{\text{B}}
  Td}{\hbar c} \right)^2 \right] .  \label{eq:tautherm3}
\end{eqnarray}
Within this approximation the kernel (\ref{eq:hdef}) for the blurred
interference pattern attains a particularly simple form. It turns into a
Gaussian of unit area whose width $\sigma_s$ is given by of $\sigma^2_s = 2 m
{\mathcal{A}_{\varepsilon}  \left( k_{\text{B}} T \right)^5 \left( L / \hbar
p_z \right)^3} / \left( 3 \pi^2 c^4 \right)$.

It should be emphasized that the decoherence time $\tau_{\tmop{th}}$ is
expedient only with respect to the particular interferometric context. Given a
choice for the time of flight $L / v_z$ and the grating separation $d$, it
determines the temperature beyond which the wave nature of the particle ceases
to be observable.

Note also that decoherence was considered to occur only between the double
slit and the detector, so far. In the same manner one may incorporate
decoherence events taking place {\tmem{in front}} of the double slit, although
this depends on the details of how the transverse coherence is produced. It is
straightforward if a single, narrow ``coherence slit'' is placed in front of
the double slit. The effect is then given by a second convolution of the form
(\ref{eq:Idec1}), with $L$ in the kernel replaced by the distance between
coherence slit and double slit. If those distances are equal one just has to
multiply the right-hand side of Eqs.~(\ref{eq:tautherm1}),
(\ref{eq:tautherm2}), and (\ref{eq:tautherm3}) by the factor of 2.   

\begin{figure}[ptb]
\includegraphics[width=\columnwidth]{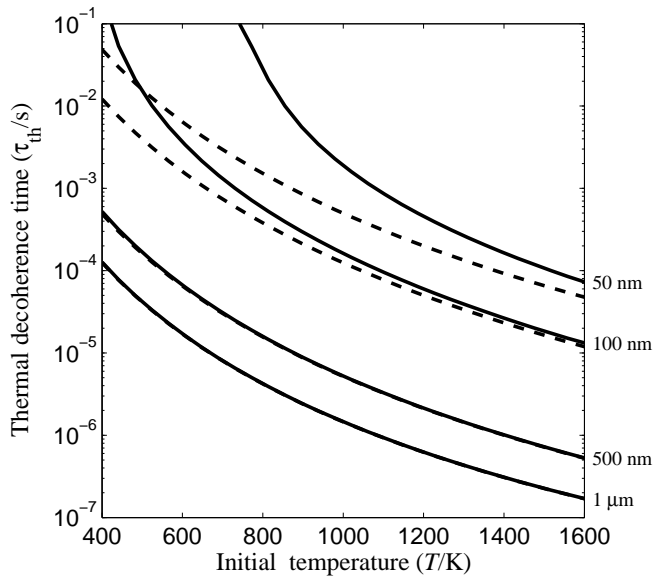}
\caption{Thermal decoherence time for double slit interference as a
  function of the initial particle temperature. The curves correspond
  to a slit separation $d$ of 50\,nm, 100\,nm, 500\,nm, and 1\,$\mu$m 
(top to bottom). The solid lines incorporate the effect of
  cooling as described by (\ref{eq:Vth}) and (\ref{eq:Rarea}), while
  the dashed lines show the result (\ref{eq:tautherm2}) for an
  asymptotically large heat capacity (they are indistinguishable for
  $d = 500\,\tmop{nm}, 1\,\mu m$). The particle parameters
  $\mathcal{A}_{\varepsilon} = 5 \times 10^{- 18} \text{m}^2$ and $C_V
  = 12000 k_{\text{B}}$ correspond to ultrafine carbonaceous
  aerosols, as discussed in the text.}
\end{figure}

Figure 1 shows the decoherence time $\tau_{\tmop{th}}$, i.e., the time of
flight where the visibility reduces to $\mathcal{V} = \mathe^{- 1}$ as
calculated by a numerical inversion of (\ref{eq:Vth}), together with the
limiting result (\ref{eq:tautherm2}) (dashed lines). They are given as a
function of the initial temperature and for several slit separations $d$. Our
choice of the two particle parameters, the effective surface area
$\mathcal{A}_{\varepsilon}$ and the heat capacity $C_V$, is given in the
caption. These values correspond to ultra-fine carbonaceous aerosols. They
were obtained by assuming a particle mass of 10$^5$\,amu, a density of
10$^3$\,kg/m$^3$, a specific heat capacity of 10$^3$\,J/(kg\,K), and a
mass specific absorption cross section of 7.5$\times$10$^3$\,m$^2$/kg as
reported in {\cite{Bond2006a}}.

One observes in Fig.~1 that the decoherence times depend strongly on
the slit separation and on the temperature. This is due to the fact
that a thermal photon emission can be harmful to interference only if
its wave length is sufficiently small to (partially) resolve the path
separation by carrying which-path information to the environment. The
rate of these harmful events decreases with decreasing temperature,
which raises the permissible time of flight, i.e., the de Broglie wave
length of the particle beam. The difference between the solid lines
and dashed lines in Fig.~1 shows that the effect of cooling can be
relevant even for the considered mesoscopic particles. In particular
for small slit separations, which are much more easily coherently
illuminated, the harmful photons carry so much energy that the
probability of further harmful emissions is substantially diminished.
This leads to a rather strong increase of $\tau_{\tmop{th}}$ with
respect to the limiting result (the dashed lines). Only for
separations around 500 \, nm the two descriptions start to be
indistinguishable.

It must be emphasized that the above results for mesoscopic particles do
{\tmem{not}} apply to large molecules, such as fullerenes, even if they
display a continuous thermal radiation spectrum. This can be seen in Fig.~2
which shows the decoherence times for a beam of C$_{70}$ fullerene molecules.
The data was obtained by numerically integrating the cooling equation and
inverting (\ref{eq:Vth}), using empirical data for the frequency-dependent
absorption cross section, as discussed in {\cite{Hornberger2005a}}. One
observes that considerably larger temperatures are required to induce
decoherence times comparable to those of Fig.~1. More importantly, the
mesoscopic result (\ref{eq:tautherm2}), indicated by the dotted lines, cannot
yet be applied, not even for a qualitative description of the curves.

\begin{figure}
\includegraphics[width=\columnwidth]{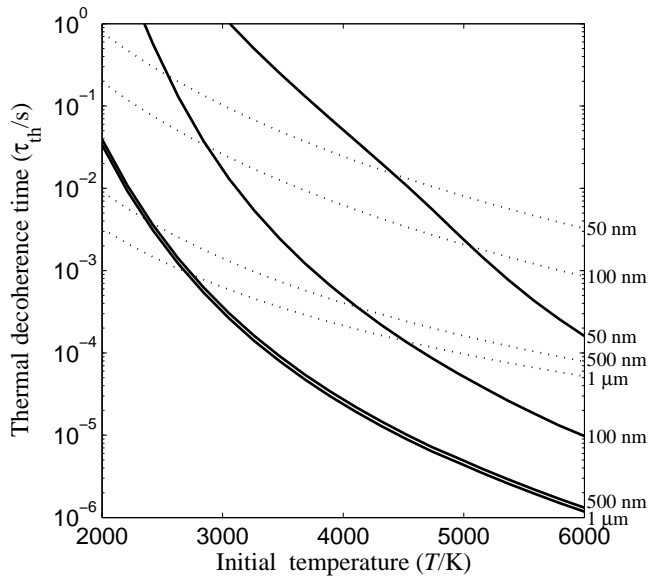}
\caption{Thermal decoherence time for double slit interference with
  fullerene molecules. The solid lines were calculated from
  Eq.~(\ref{eq:Vth}) using the full frequency- and cooling-dependent
  emission rate. Compared to Fig.~1, similar decoherence times are
  found only at considerably larger temperatures. The comparison with
  the dotted lines shows that for fullerenes the mesoscopic result
  (\ref{eq:tautherm2}) is not yet applicable, not even for a
  qualitative description of the temperature dependence of the
  interference contrast ($\mathcal{A}_{\varepsilon} = 10^{-
    22}$\,m$^2$  {\cite{Kolodney1995b}}).}
\end{figure}

It is remarkable that the slit separations of $d = 500 \, \tmop{nm}$
and of $d = 1 \, \mu m$ display an almost equal dependence on thermal
decoherence. This behavior is due to the gap in the electronic
excitation spectrum of fullerenes, which strongly suppresses radiation
above about $800 \, \tmop{nm}$. As a consequence, interference with
slit separations beyond that length is affected similarly by the
emitted photons. It should also be emphasized that the effect of
cooling is much stronger than in the case of Fig.~1. Its neglect
would render the decoherence times larger by an order of magnitude for
$d = 50 \, \tmop{nm}$, and still about threefold larger for $d = 1 \,
\mu m$ (not shown). This is part of the reason why the present results
differ markedly from those of Ref.~{\cite{FacchiNonsense}}, which
account neither for the effect of cooling nor for the detailed
emission spectrum of fullerenes
{\footnote{The resulting series
    expansion given in Ref.~{\cite{FacchiNonsense}}(b) diverges. For
    the comparison we used the pseudo-converged first few terms, as
    was apparently done for producing the figures given there.}}.

\section{Conclusions}

The present article described the effect of momentum-exchange mediated
decoherence on far-field interference of material waves. Rather than solving
the dynamic equation of motion, a phase space formulation for the stationary
scattering problem was employed. This way exact and transparent expressions
could be derived for the modification of the interference pattern due to the
presence of decoherence.  Although we specialized to the case of double slit
interference, the generalization to arbitrary multi-slit arrangements is
straightforward in the present formulation.

We calculated the reduction of the interference visibility due to the
particle's heat radiation and found that a realistic description of molecular
objects requires accounting for the effect of cooling and for the particle
color, i.e., the deviations from the blackbody behavior. A particularly
simple scaling behavior could be obtained for mesoscopic particles, which are
characterized only by their emissivity and heat capacity.

As a final point, let us illustrate the temperature requirements for an object
of the size of a small virus, assuming a mass of $5 \times 10^7 \,
\tmop{amu}$. The rivaling effect of {\tmem{collisional decoherence}} was
estimated for such a particle to limit the permissible pressure of background
gases to $p_0 = 2.7 \times 10^{- 11} \, \tmop{mbar} \left( \tau / \text{s}
\right)^{- 1}$ {\cite{Colldeco}}. Using the above-mentioned mass-specific
absorption cross section {\cite{Bond2006a}} we find from (\ref{eq:tautherm3})
that the characteristic temperature is as low as $T_0 = {19 \, \text{K} \,
\left( d^2 \tau / \mu \text{m}^2 \text{s} \right)^{- 1 / 5}}$. Inserting
reasonable values for the slit separation and the time of flight, $d = 0.5 \,
\mu \text{m}$ and $\tau = 0.1 \, \text{s}$, this yields about 40 \, K. It
indicates that the thermal decoherence may be more difficult to control than
the effect of background gases for particles of that size. It must be
emphasized, however, that in the foreseeable future interferometry of
mesoscopic objects will be limited by the practical difficulty of producing
brilliant particle beams and of avoiding vibrational noise
{\cite{Stibor2005a}}, rather than by the ``natural'' decoherence mechanism of
heat radiation.

{\subsection*{Acknowledgments}}

Helpful discussions with Markus Arndt are gratefully acknowledged. This work
was supported by the DFG Emmy Noether program.

\end{document}